# Aharonov-Borm oscillation and Microwave-induced edge-magnetoplasmon modes enhanced by quantum point contact

Xinghao Wang*, Wenfeng Zhang

International Center for Quantum Materials, School of Physics, Peking University, Beijing 100871, China

L. N. Pfeiffer, K. W. Baldwin, and K. W. West

Department of Electrical Engineering, Princeton University, Princeton, NJ 08544, USA

## Abstract

AB oscillation in weak magnetic field ($B < 1.5kG$) is observed in QPC due to interference between electrons propagating along different QPC channels. We also investigate photo-induced magnetoresistance oscillation in open-regime split-gate QPC under MW irradiation. It is attributed to EMPs interfering in the QPC region. The influence of MW power, frequency and split gate voltage is discussed thoroughly. We unify the result of photoconductance at $B = 0$ with EMP theories.

* 17wxhwypku@pku.edu.cn

*Introduction.* - In low magnetic field $B$ together with microwave (MW) irradiation, a large quantity of interesting condensed matter physics has been extensively studied in two-dimensional electron gas (2DEG). Cyclotron resonance $\omega_c = \omega$ is the first to be paid attention to and it clearly demonstrates Landau level (LL) spectra. Here, $\omega$ is MW frequency and $\omega_c = eB/m^*$ is cyclotron frequency and $m^*$ is effective electron mass. Two decades ago, MW-induced resistance oscillation (MIRO) [1, 2] and accompanied zero-resistance state (ZRS) [3, 4] were discovered. This $1/B$-periodic oscillation is attributed to photon-assisted impurity scattering of electrons between LLs. These phenomena obey the famous Kohn's theorem [5] that in 2DEG with translation symmetry photons do not influence the relative motion of carriers but only couple to the center-of-mass motion. If translation symmetry is broken, collective modes that do not follow Kohn's theorem can be detected, e. g., finite-wave-vector magnetoplasmons (MPs) induced by grating couplers [6] or limited width of Hall bar [7], giant second harmonic peak due to transverse magnetosonic waves [8, 9] and edge magnetoplasmons (EMPs) with $B$-periodic resistance oscillation [10-12].

EMP is a special collective mode of plasmon propagating along the sample edge. It has very wide frequency range from $kHz$ to $THz$ and small damping in contrast to the bulk plasmon. EMPs were used to be studied through plasmon resonances by adding ac bias between contacts [13-18] until it was discovered that interference among EMPs generated at different contacts by MW irradiation can bring about $B$-periodic resistance oscillation [10-12]. As is predicted by theories [19, 20], resistance influenced by EMPs is proportional to $|1 + \exp(iqL)|^2$. Here EMP wave vector $q \propto \omega B/n$, $L$ is difference of plasmon propagation length and $n$ is carrier density. Compared with other low temperature physics, EMPs are extremely robust against thermal excitation and the resistance oscillation exists even at $80K$. There are debates on whether $L$ is the distance between contacts measured along sample edge or perimeter of 2DEG due to chiral propagation path of electrons. High-mobility samples ($\mu \sim 2 \times 10^6 cm^2/Vs$) are in favor of the former picture [12] while ultrahigh-mobility ones ($\mu \sim 1 \times 10^7 cm^2/Vs$) consists with the latter [10-11].

EMPs generated by MW irradiation can be greatly magnified in narrow constriction called quantum point contact (QPC) [21, 22] due to enhancement of interference. QPC has another advantage that MIRO is prohibited by narrow constrictions so that MIRO and EMP can be easily distinguished. Giant microwave-induced $B$-periodic resistance oscillation with a relative amplitude of about 700% was reported in a bridged-gate tunnel point contact [23].

Despite photoconductance of QPC is remarkable and significant, there is no EMP experiment of QPC in open regime where electrons move ballistically through the channel. In this paper, we report transport measurement results of split-gate QPC under MW irradiation in several high-mobility or ultrahigh-mobility samples. Incidentally, we observed Aharonov-Borm (AB) effect with good-quality resistance oscillation in our high-mobility QPCs without MW irradiation.

*Experimental set up.* - Our experiment is performed in a ³He refrigerator (base temperature $T_b = 0.3K$) with two wafers. After being illuminated by red light-emitting diode at $2K$, one is high-mobility 2DEG with $\mu$ around $2 \times 10^6 cm^2/Vs$ and the other has better quality with $\mu$ above $2 \times 10^7 cm^2/Vs$ at 0.3 K. Carrier density $n$ is about $2.5 \times 10^{11} cm^{-2}$ for both two wafers. Split-gate QPC is defined by e-beam lithography and Ti/Au top gate. Electrical contacts are made by Ge/Pd/Au alloy annealed at $450°C$. In our experiment, MW with frequency $f$ ranging from $26.5 GHz$ to $90 GHz$ is generated by Gunn oscillators and MW power $P_{MW}$ is adjusted by a programmable rotary vane attenuator. The attenuation of MW power is $P_{MW}/P_0$, where $P_0$ is the original power of Gunn oscillator. We measure the longitudinal resistance $R_L$ through each QPC (Fig.1(a)) with low-frequency lock-in technique.

*Aharonov-Borm effect in QPC.* - Before demonstrating the results of EMP-induced resistance oscillation in QPC, we are going to show the good quality of our QPC samples in dark environment. In QPC, electrons propagate ballistically in quasi-one-dimensional electron gas channels. QPC resistance $R_L = h/2N_{QPC}e^2$ is decided by number of quasi-one-dimensional edge modes $N_{QPC}$. Quantization of QPC resistance indicates that QPC is in good quality, which differs from common point contact used in previous EMP experiment [23-25]. As one of our samples, QPC A from the high-mobility wafer has rigorously quantized plateaus for $N_{QPC} = 1 - 5$, and there are developing plateaus for $N_{QPC} = 6 - 12$ (Fig.1(b)). Other samples in our experiment show similar results.

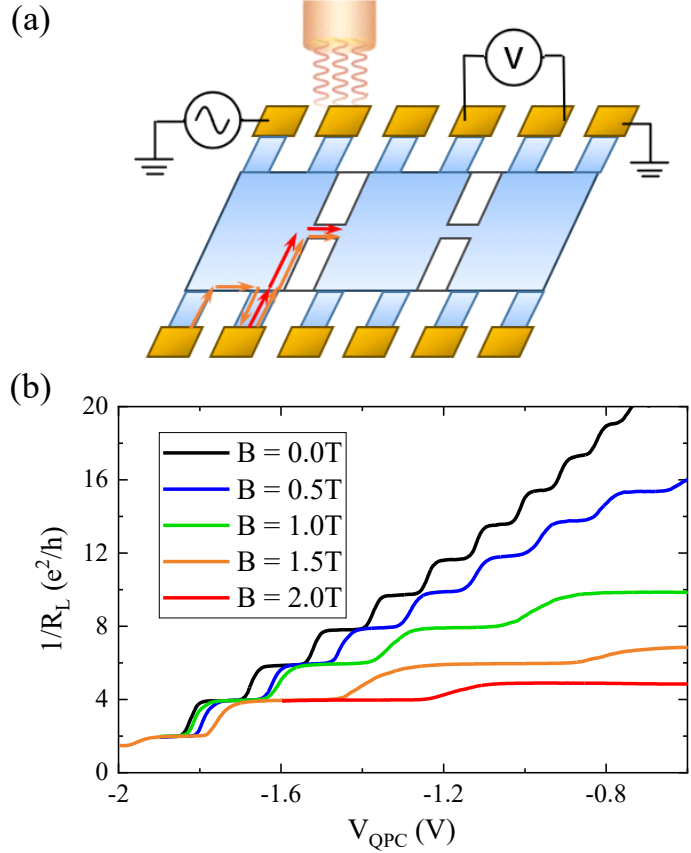

Fig. 1 (a) Under MW irradiation, longitudinal resistance across a QPC is measured through lock-in amplifier (voltmeter). The principle behind EMPs-induced resistance oscillation is shown in this diagram. Two EMPs (red and orange) originate from neighbouring contacts and interfere at the QPC region, which result in B-related resistance oscillation. (b) In the dark environment, QPC conductance plateaus at $2N_{QPC}e^2/h$ have good quality with/without magnetic field. The plateaus at $(2N + 1)e^2/h$ are induced by integer quantum Hall effect but not spin-splitting of QPC band.

The good quality of our samples can also be proved through another phenomenon accidentally discovered in the measurement. In Fig.2(a), $R_L$ oscillates with magnetic field $B$ periodically without MW irradiation, which is presented more clearly in $d^2R_L/dB^2$. The period is $\Delta B = 0.50 kG$. This is concrete evidence of AB effect in QPC. When an electron in 2DEG approaches the QPC region in weak magnetic field, it chooses different conducting channels in order to move through the QPC (Fig.2(d)). After traversing the QPC, electrons in different channels interfere with each other. AB phase is accumulated in the process and results in AB oscillation.

According to the oscillation period, the effective area that two paths of electrons encircle is $h/e\Delta B = 0.0827 \mu m^2$, which is about half the size of QPC region $0.4 * 0.4 = 0.16 \mu m^2$. This result makes sense since effective area surrounded by ballistic channels is averaged and thus smaller than the actual QPC area. For QPC B with area equal to $0.4 * 1.0 = 0.40 \mu m^2$, period of AB oscillation is $\Delta B = 0.25 kG$, its amplitude is larger (Fig.2(b)). and its periodicity is shown in Fig.2(e). The size-dependent period conforms with AB effect. The AB oscillation of QPC B becomes too weak to be observed when $B > 1.5 kG$ probably for the reason that electrons, with cyclotron radius $R_c < 0.5 \mu m$, can move in circles in the QPC region (QPC width $\sim 1 \mu m$). Circulating electrons cannot interfere with each other since their propagation length turns decoherent after several circulations. This makes AB effect in QPC distinguished from ordinary ring interferometer whose oscillation persists till several Tesla.

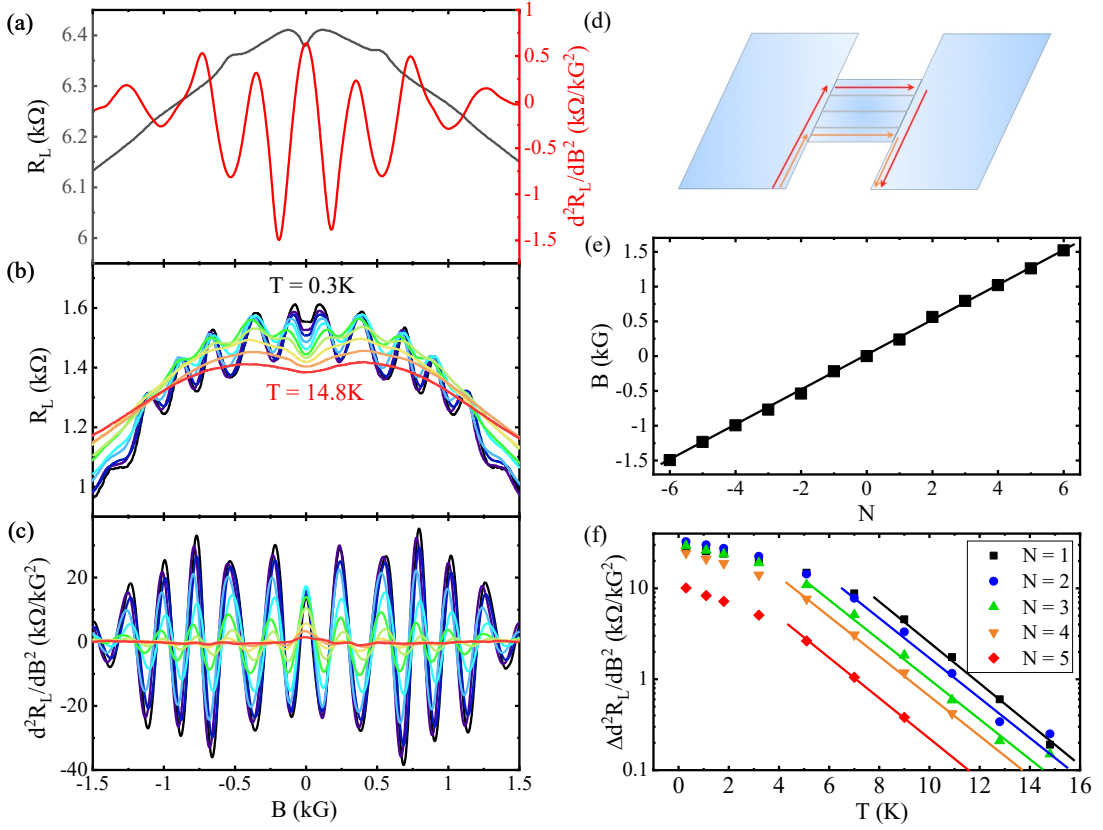

Fig. 2 (a) We found out a QPC-related resistance oscillation that is linear with $B$ in a 0.4$um$ QPC. The oscillation disappears at high magnetic field $B > 1.5kG$. (b, c) The oscillation is stronger in a 1$um$ QPC and smears out at $T = 14.8K$. Temperature dependence of resistance oscillation is plot in the two subfigures. (d) A simple interpretation for this peculiar resistance oscillation is AB oscillation and its phase difference comes from distinct conductance channels in the QPC. (e) The oscillation is periodic in $B$. (f) At high temperature, $R_L$ is proportional to $\exp(-\alpha T)$, indicating that decoherence of AB oscillation happens. At low temperature, $\Delta R_L$ is limited by thermal smearing of ballistic transport and disorder-induced decoherence.

We also measured temperature dependence of the AB oscillation (Fig.2(b, c)) and temperature dependence of oscillation amplitude $\Delta R_L \propto \Delta d^2 R_L / dB^2$ is shown in Fig.2(f). As for ordinary AB effect discovered in normal metal [26] or ring interferometer [27], there are two mechanisms that determine the temperature dependence of AB oscillations, (i) thermal smearing of ballistic transport and (ii) decoherence of electron waves. For mechanism (i), $\Delta R_L$ is proportional to $T^{-1/2}$, which shares similarities with the thermal averaging of conventional fluctuation of conductance. For mechanism (ii), $\Delta R_L \propto \exp(-l/L_\phi)$ and propagation length $l$ larger than coherence length $L_\phi$ ($L_\phi \propto T^{-1}$) results in decoherence of AB oscillations. As shown in the figure, temperature dependence of our QPC sample at high temperature unambiguously follows mechanism (ii). This makes sense since propagation length $l \sim 1\mu m$ is quite large for AB oscillations. However, at low temperature, $\Delta R_L$ is limited by other mechanisms including thermal smearing of ballistic transport and disorder-induced decoherence.

To the authors' knowledge, this is the first time to report the existence of AB oscillation in QPC. This fact vividly illustrates the good quality of our samples and empirically speaking, QPCs in high-mobility ($\mu \sim 10^6 cm^2/Vs$) samples tends to work better than those of ultrahigh-mobility ($\mu \sim 10^7 cm^2/Vs$). Considering that EMP-induced resistance oscillation and AB oscillation are both $B$-periodic, we are going to compare these two effects in the following part of this article.

*Resistance oscillation under MW irradiation.* – Unlike AB oscillation, MW induced $R_L$ oscillation is much more complex. EMP-induced resistance oscillation is usually asymmetric on zero magnetic field. Another main feature of this oscillation is that MW power and sample geometry could significantly, or even qualitatively influence the result. Due to complexity of the experimental result, we emphasize that good quality of our QPC sample helps us understand it.

Compared with QPCs in tunnelling regime, QPCs in open regime show relatively weak $B$-periodic resistance oscillation with $\Delta R_L/R_L < 1$ when $\omega_c > \omega$ but strong resistance oscillation when $\omega_c < \omega$ (Fig.3(a)). The former is consistent with EMPs studied in [10, 11, 23] since its frequency dependence shown in Fig.3(b) repeats the result that $\Delta B$ is reciprocal to MW frequency $f$. The latter shows resistance oscillation with $\Delta R/R > 1$ when MW power is strong enough. Although clear segmentation of longitudinal resistance around $\omega_c = \omega$ is similar to the case in reference [23], low-magnetic-field regime ($\omega_c < \omega$) was neglected. In our samples, the huge photo-induced resistance at $\omega_c < \omega$ sensitively depends on MW power $P_{MW}$ and clearly overweighs the normal EMP-induced resistance oscillation in high magnetic field. As MW power gradually increases, resistance oscillations at $\omega_c < \omega$ prevails earlier than the normal ones at $\omega_c > \omega$ (Fig.3(c)). In low-magnetic-field regime, there are some magnetic fields where photo-induced resistance is relatively weak (marked in Fig.3(a)). These positions, e.g., the resistance minimum at $B = \pm 1.25 kG$, are attributed to destructive interference of EMP, since resistance minimum at both $\omega_c < \omega$ and $\omega_c > \omega$ shows the same periodicity (inset of Fig.3(a)).

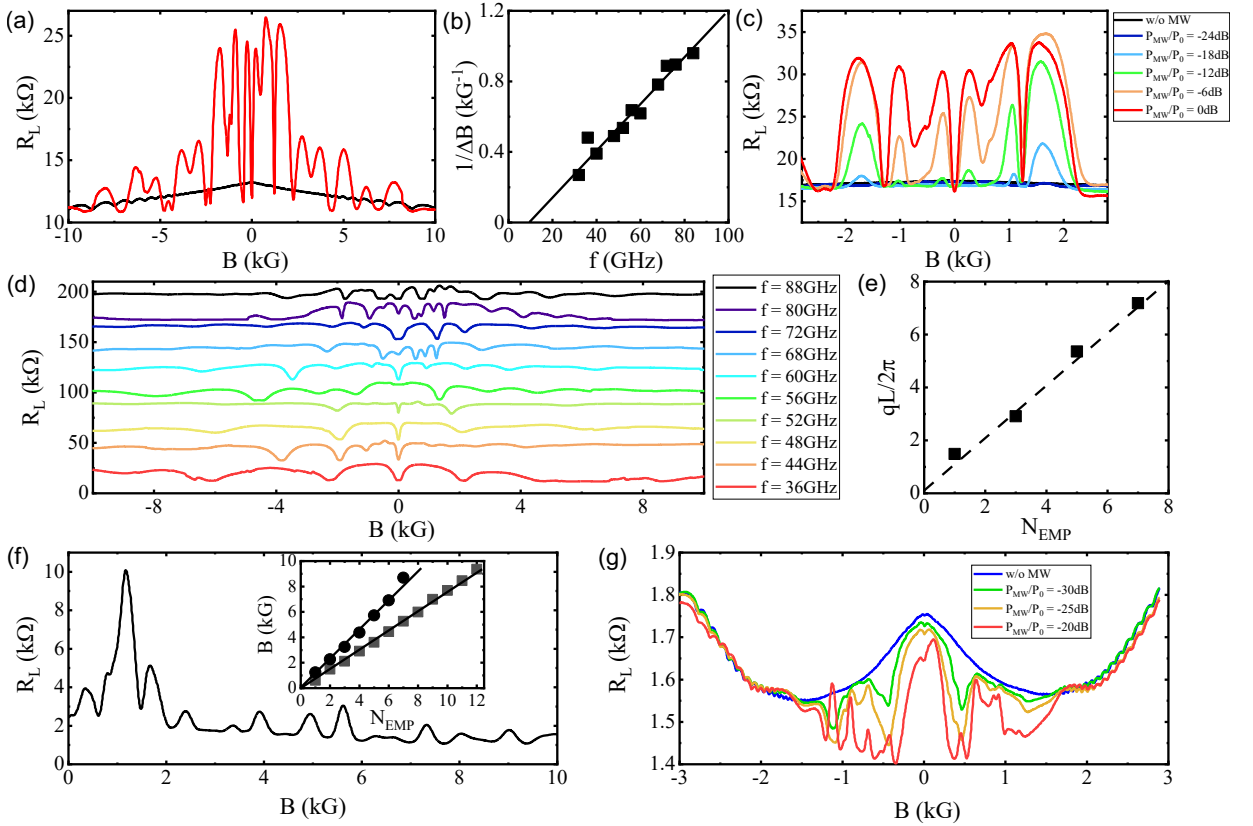

Fig.3 (a) Negative photoconductance is induced by MW irradiation for QPCs in open regime. The trace shows relatively weak $B$-periodic resistance oscillation when $\omega_c > \omega$ but strong resistance oscillation when $\omega_c < \omega$. (b) The $B$-periodic resistance oscillation follows the same feature of EMPs: $\Delta B \propto 1/f$. (c) The saturation of photoconductance under strong enough MW power vividly demonstrates how EMP-induced resistance oscillation evolves when MW power varies. (d) EMP-induced resistance oscillation relies on MW frequency $f$ (For clarity, data for different MW frequency is consecutively shifted upward by $20k\Omega$.), (e) and disappearing of the dip at zero magnetic field shows the periodicity of $qL = 2\pi N_{EMP}$, in which way we can measure propagation length difference $L$. (f) For an ultrahigh-mobility QPC ($\mu \sim$ $2 \times 10^7 cm^2/Vs$) but similar density ($n = 2.65 \times 10^{11} cm^{-2}$) under MW radiation with the same frequency, $B$-period of EMPs is about half of that of high-mobility samples. The inset shows the period of the high- (circles) and ultrahigh- (squares) mobility samples. (g) When MW irradiation is weak and sample mobility is extremely high, the model of photo-induced energy barrier does not dominate. The resistance peaks induced by EMPs turn into resistance dips due to superballistic transport of electrons.

There is another difference between our result and that of reference [23]. In tunnelling regime ($G_{QPC} < e^2/2h$), photoconductance is positive due to photon-assisted tunneling. Contrarily, in open regime QPC ($G_{QPC} > e^2/2h$) conductance $G_{QPC}$ decreases under MW irradiation, which is consistent with the bridged-gate result without magnetic field in [24]. MW irradiation influences the electron distribution function near the constrictions [25], which in turn effectively modifies the energy barrier of QPC, $U_0$, to $U_0 + A cos(\omega t)$. The broadening energy barrier together with Fermi energy $E_F$ can essentially affect the result of MW irradiation. Due to the form of electron distribution function $F(E)$ and its delta-function like derivative $\partial F(E)/\partial E$, energy barrier that is close to $E_F$ makes greater influence on conductance $\sigma = -\sigma_0 \Sigma \int dE T(E) \partial F(E)/\partial E$, where $\sigma_0 = 2e^2/h$ and $T(E)$ is the transmission coefficient. This means that in tunneling regime MW effectively reduces energy barrier (from $U_0$ to $U_0 - \delta A$), while in open regime MW enlarges energy barrier (from $U_0$ to $U_0 + \delta A$). Detailed computational and experimental results can be referred to reference [24, 25].

The picture illustrated above is also consistent with the saturation of photoconductance under strong enough MW power (Fig.3(c)). In the regime of $\omega_c < \omega$, the outer peak is much sensitive to MW power and relatively weak MW ($P_{MW}/P_0 = -18dB$) can bring about distinguished resistance peaks. As MW power gradually increases, the resistance peaks saturate to a value of about twice the original QPC resistance. This is because QPC resistance with MW irradiation is approaching the crossover value between open and tunneling regime, $2h/e^2$. Saturation of photoconductance can be understood in the following picture. When MW power is strong enough so that $|E_F - U_0| < A$, the effective barrier height $U_0$ approaches $E_F$ and stops increasing, which results in irradiated QPC resistance $R_L \sim h/e^2$.

In this way, we can observe a saturation pattern of QPC magnetoresistance in Fig.3(d). In these traces, there are some magnetic fields where MW irradiation hardly influences $R_L$. These minimums are exactly attributed to the destructive interference of EMPs in the QPC region although some of them have relatively large negative photoconductance due to incoherence of EMPs. This process can help us understand physics of EMPs further. At resistance minimums, destructive interference of EMPs protects QPC resistance from influence of MW irradiation, for there is no EMP in the QPC region and energy barrier $U_0$ is unchanged. We can then unify the theories of EMPs [19, 20] together with those of photo response of QPC at zero magnetic field [24, 25].

At zero magnetic field, there is always a resistance dip apart from some certain MW frequencies, since magnetoresistance is an approximate even function of $B$. Whether there is a resistance dip is decided by the phase difference $qL$ of two interfering EMPs. When EMP wave vector $q = N_{EMP}\pi/L$ at $B = 0$ and $N_{EMP}$ is an odd integer, a resistance minimum appears, e.g., the trace of $f = 72GHz$. While there is almost no sign of resistance minimum at $B = 0$ in the trace of $f = 56GHz$ because constructive interference happens. Since energy dispersion of EMP is $\omega_p^2 = ne^2q/(\epsilon_{GaAs} + 1)\epsilon_0 m^*$ and $\omega_p^2 = 1.217\omega^2$ at $B = 0$ (according to reference [28]), we know that phase difference $qL \propto f^2$. By changing MW frequency $f$, resistance dip at $B = 0$ disappears and reappears for several times. In this way, we can decide propagating length difference $L$ of EMP in our sample. The fitted result is $L = 540\mu m$, which is about the distance between two contacts which is about $580\mu m$ (Fig.3(e)). This result is deduced in a brand-new method and also in favor of reference [17]. $L$ equals $560\mu m$ by computing the EMPs period which is used in reference [23], indicating the consistency of the EMP theories.

For a QPC with ultrahigh mobility but similar density under MW radiation with the same frequency, $B$-period of EMPs is about half of that of high-mobility samples (Fig.3(f)). This indicates that for ultrahigh-mobility samples, $L$ is doubled because of longer correlation length of EMPs when impurity scattering decreases. Since $L$ is much shorter than perimeter of our samples, we rule out the possibility of EMP travelling around the sample edge and interfering at the same place even though the ultrahigh-mobility samples have better quality than those in reference [12].

Under relatively weak MW irradiation, photoconductance shows more complicated resistance oscillation in low-magnetic-field regime ($\omega_c < \omega$) (Fig.3(c)). Whether the resistance peak is strong or weak is decided by MW absorption rate at certain magnetic field. As a result, magnetoresistance of a QPC under MW irradiation can vividly demonstrate its MW absorption spectrum. We notice that in Fig.3(c), the resistance peak near resonance absorption ($B = 1.7kG$) is reasonably much more sensitive to MW irradiation. This effect is interpreted as resonance absorption of MW irradiation near $\omega_c = \omega$. As MW power enhances, the resistance peak becomes higher and broader, and is then truncated by the resistance minimum at $B = 1.25kG$, i.e., the two peaks beside $B = 1.25kG$ can be viewed as one peak split by destructive interference of EMPs. The photo-resistance is truncated by a weaker minimum at $B = 0.68kG$ due to double of contact distance $L$ between second nearest contacts.

If MW irradiation is even weaker and the model of photo-induced energy barrier does not dominate (Fig.3(g)), we can observe an intriguing effect in our samples. The resistance peaks we discuss above turns into resistance dips in the same magnetic field. This does not mean there is a $\pi$ phase under weak MW irradiation or the picture of QPC in open regime does not conform to the experimental results. We are not going to discuss this effect in this manuscript and it is illustrated in reference [29] as superballistic flow of electrons in hydrodynamic 2DEG systems [30].

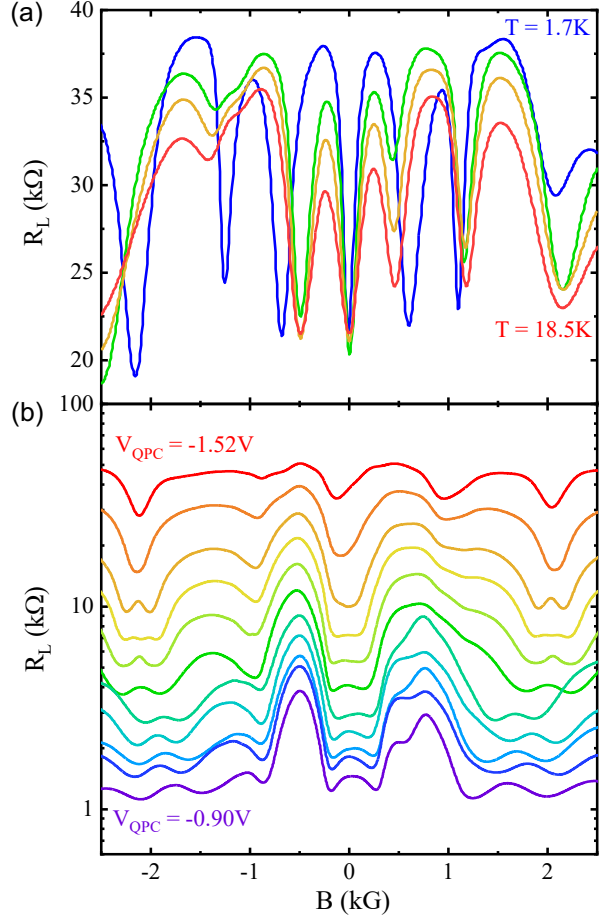

Fig.4 (a) Temperature-dependence of EMP-induced resistance oscillation indicates that EMPs are robust against heating. (b) By reducing QPC conductance, MW absorption rate of QPC increases and photoconductance becomes saturate.

Temperature-dependence of EMP-induced resistance oscillation is shown in Fig.4(a). The photo-induced resistance is insensitive to temperature, which indicates that EMPs are robust against heating. The resistance oscillation persists until several tens of Kelvin because EMP can always propagate if $\omega\tau > 1$, where $\tau$ is transport relaxation time [20]. As for high-mobility GaAs/AlGaAs heterostructures and MW frequency around $71 GHz$, this condition is easily satisfied below liquid nitrogen temperature. Temperature robustness of the resistance oscillation also hints that EMP can tolerate relatively high electron temperature in 2DEG.

We have also measured EMP's response to QPC gate voltage $V_{QPC}$ (Fig.4(b)). Changing $V_{QPC}$ is equivalent to adjusting energy barrier $U_0$ of the narrow constriction. We find that as $U_0$ increases ($V_{QPC}$ decreases), the photoconductance saturates more and more easily, as if MW power is enhanced. We attribute this effect to increasing MW absorption rate of QPC induced by decreasing QPC conductance, which has been discussed in paper [31]. By increasing $U_0$, EMP-induced ac energy barrier $A$ also increases, which results in greater photoconductance.

Photo-induced magnetoresistance oscillation in QPCs influenced by interference among EMPs has both pros and cons. On one hand, split gate QPCs in open regime can also work as a sensitive MW detector to measure MW frequency and power just like bridged-gate ones in tunneling regime. Under MW irradiation with enough power, the resistance minimum of destructive interference of EMPs is always sharp. On the other hand, this effect seems to forbid us to use quantum-dot thermometry [32] to measure electron temperature in 2DEG under MW irradiation at low temperature. However, since MW irradiation easily warms up electrons in QPC, such thermometry is not necessary and we can use other methods to measure electron temperature.

*Conclusion.* - AB oscillation in weak magnetic field ($B < 1.5kG$) is observed in QPC due to interference between electrons propagating along different QPC channels. We also investigate EMP-induced magnetoresistance oscillation in open-regime split-gate QPC under MW irradiation. The influence of MW power, frequency and split gate voltage is discussed thoroughly. We unify the results of photoconductance at $B = 0$ with EMP theories by measuring EMP propagation length in a new method. EMP-induced resistance oscillation is compatible with negative photoconductance of QPC in the open regime.

*Acknowledgements.* The work at PKU was funded by the National Key R&D Program of China (Grants No. 2017YFA0303300 and 2019YFA0308400), by the Strategic Priority Research Program of Chinese Academy of Sciences (Grant No. XDB28000000). The work at Princeton was funded by the Gordon and Betty Moore

Foundation through the EPiQS initiative Grant No. GBMF4420, by the National Science Foundation MRSEC Grant No. DMR-1420541.

**References**

[1] M. A. Zudov, R. R. Du, J. A. Simmons, and J. L. Reno, "Shubnikov–de Haas-like oscillations in millimeterwave photoconductivity in a high-mobility two-dimensional electron gas", *Phys. Rev. B* **64**, 201311(R) (2001).

[2] P. D. Ye, L. W. Engel, D. C. Tsui, J. A. Simmons, J. R. Wendt, G. A. Vawter, and J. L. Reno, "Giant microwave photoresistance of two-dimensional electron gas", *Appl. Phys. Lett.* **79**, 2193 (2001).

[3] R. G. Mani, J. H. Smet, K. von Klitzing, V. Narayanamurti, W. B. Johnson, and V. Umansky, "Zero-resistance states induced by electromagnetic-wave excitation in GaAs/AlGaAs heterostructures", *Nature* **420**, 646 (2002).

[4] M. A. Zudov, R. R. Du, L. N. Pfeiffer, and K. W. West, "Evidence for a New Dissipationless Effect in 2D Electronic Transport", *Phys. Rev. Lett.* **90**, 046807 (2003).

[5] W. Kohn, "Cyclotron Resonance and de Haas-van Alphen Oscillations of an Interacting Electron Gas", *Phys. Rev.* **123**, 1242 (1961)

[6] S.J. Allen, Jr., D.C. Tsui, and R.A. Logan, "Observation of the Two-Dimensional Plasmon in Silicon Inversion Layers", *Phys. Rev. Lett.* **38**, 980 (1977).

[7] E. Vasiliadou, G. Muller, D. Heitmann, D. Weiss, and K. v. Klitzing, H. Nickel, W. Schlapp, and R. Losch, "Collective response in the microwave photoconductivity of Hall bar structures", *Phys. Rev. B* **48**, 17145 (1993).

[8] Y. Dai, R. R. Du, L. N. Pfeiffer, and K. W. West, "Observation of a cyclotron harmonic spike in microwave-induced resistances in ultraclean Ga17As/AlGaAs quantum wells", *Phys. Rev. Lett.* **105**, 246802 (2010).

[9] P. S. Alekseev and A. P. Alekseeva, "Transverse magnetosonic waves and viscoelastic resonance in a two-dimensional highly viscous electron fluid", *Phys. Rev. Lett.* **123**, 236801 (2019).

[10] I. V. Kukushkin, et. al., "New Type of B-Periodic Magneto-Oscillations in a Two-Dimensional Electron System Induced by Microwave Irradiation", *Phys. Rev. Lett.* **92**, 236803 (2004).

[11] I. V. Kukushkin, et. al., "Miniature quantum-well microwave spectrometer operating at liquid-nitrogen temperatures", *Appl. Phys. Lett.* **86**, 044101 (2005).

[12] K. Stone, et.al., "Photovoltaic oscillations due to edge-magnetoplasmon modes in a very high-mobility two-dimensional electron gas", *Phys. Rev. B* **76**, 153306 (2007).

[13] S. J. Allen, H. L. Storrner, and J. C. M. Hwang, "Dimensional resonance of the two-dimensional electron gas in selectively doped GaAs/A1GaAs heterostructures", *Phys. Rev. B* **28**, 4875 (1983).

[14] D. B. Mast, A. J. Dahm, and A. L. Fetter, "Observation of Bulk and Edge Magnetoplasmons in a Two-Dimensional Electron Fluid", *Phys. Rev. Lett.* **54**, 1706 (1985).

[15] T. Demel, D. Heitrnann, P. Grambow, and K. Ploog, "Nonlocal Dynamic Response and Level Crossings in Quantum-Dot Structures", *Phys. Rev. Lett.* **64**, 788 (1990).

[16] M. Wassermeier, J. Oshinowo, J. P. Kotthaus, A. H. Macdonald, C. T. Foxon, and J. J. Harris, "Edge magnetoplasmons in the fractional-quantum-Hall-effect regime", *Phys. Rev. B* **41**, 10287 (1990).

[17] I. Grodnensky, D. Heitmann, and K. von Klitzing, "Nonlocal Dispersion of Edge Magnetoplasma Excitations in a Two-Dimensional Electron System", *Phys. Rev. Lett.* **67**, 1019 (1991).

[18] R. C. Ashoori, H. L. Störmer, L. N. Pfeiffer, K. W. Baldwin, and K. West, "Edge magnetoplasmons in the time domain", *Phys. Rev. B* **45**, 3894 (1992).

[19] V. A. Volkov and S. A. Mikhailov, "Edge magnetoplasmons: low frequency weakly damped excitations in inhomogeneous two-dimensional electron systems", *Sov. Phys. JETP* **67**, 1639 (1988).

[20] S. A. Mikhailov, "Propagation of edge magnetoplasmons in semiconductor quantum-well structures", *Appl. Phys. Lett.* **89**, 042109 (2006).

[21] Y. V. Sharvin, "A possible method for studying Fermi surfaces", *Sov. Phys. JETP* **21,** 655–656 (1965).

[22] C. W. J. Beenakker & H. van Houten, "Quantum transport in semiconductor nanostructures", *Solid State Phys.* **44,** 1–228 (1991).

[23] A. D. Levin, et. al., "Giant microwave-induced *B*-periodic magnetoresistance oscillations in a two-dimensional electron gas with a bridged-gate tunnel point contact", *Phys. Rev. B* **95**, 081408(R) (2017).

[24] V. A. Tkachenko, et. al., "Low-Frequency Microwave Response of a Quantum Point Contact", *JETP Letters* **114**, 2, 110–115 (2021).

[25] A. D. Levin, et. al., "Giant microwave photo-conductance of a tunnel point contact with a bridged gate", *Appl. Phys. Lett.* **107**, 072112 (2015).

[26] S. Washburn, C. P. Umbach, R. B. Laibowitz, and R. A. Webb, "Temperature dependence of the normal-metal Aharonov-Bohm effect", *Phys. Rev. B* **32**, 4789(R) (1985).

[27] E. B. Olshanetsky, Z. D. Kvon, D. V. Sheglov, A. V. Latyshev, A. I. Toropov, and J. C. Portal, "Temperature Dependence of Aharonov–Bohm Oscillations in Small Quasi-Ballistic Interferometers", *JETP Letters* **81**, 625–628 (2005).

[28] F. Stern, "Polarizability of a Two-Dimensional Electron Gas", *Phys. Rev. Lett.* **18**, 546 (1967).

[29] ultrahigh

[30] R. K. Kumar, D. A. Bandurin, F. M. D. Pellegrino, Y. Cao, A. Principi, H. Guo, G. H. Auton, M. B. Shalom, L. A. Ponomarenko, G. Falkovich, et al., "Superballistic flow of viscous electron fluid through graphene constrictions", *Nat. Phys.* **13**, 1182 (2017).

[31] X. Wang, et. al., "Hydrodynamic charge transport in GaAs/AlGaAs ultrahigh-mobility two-dimensional electron gas", Phys. Rev. B **106**, L241302 (2022)

[32] D. Maradan, L. Casparis, T.-M. Liu, D. E. F. Biesinger, C. P. Scheller, and D. M. Zumbuhl, "GaAs Quantum Dot Thermometry Using Direct Transport and Charge Sensing", *Journal of low Temp. Phys.* **175**, 784-798 (2014).